\documentclass[floatfix,showpacs,showkeys,aps,prl,twocolumn]{revtex4-2}
\usepackage{lineno}
\usepackage{amsmath}
\usepackage{amssymb}

\usepackage{tikz}
\usepackage{lipsum,adjustbox}
\usepackage{graphicx}
\usepackage[caption = false]{subfig}
\usepackage{dcolumn}
\usepackage{bm}
\usepackage{gensymb}

\usepackage{tabstackengine}
\usepackage[inline]{enumitem}

\setstackEOL{\cr}
\setstackgap{L}{\normalbaselineskip}

\let\oldequation\equation
\let\oldendequation\endequation

\renewenvironment{equation}
  {\linenomathNonumbers\oldequation}
  {\oldendequation\endlinenomath}

\newcolumntype{L}[1]{>{\raggedright\arraybackslash}p{#1}}
\newcolumntype{C}[1]{>{\centering\arraybackslash}p{#1}}
\newcolumntype{R}[1]{>{\raggedleft\arraybackslash}p{#1}}

\usepackage{hyperref}
\hypersetup{
 pdftitle={},
 pdfauthor={},
 pdfsubject={},
 pdfkeywords={},
 pdfstartview={},
 bookmarksopen=true, breaklinks=true, debug=true,
 colorlinks=true, linkcolor=blue, citecolor=red, urlcolor=blue,
 hyperfigures=true
}

\newcommand{\BESIIIorcid}[1]{\href{https://orcid.org/#1}{\hspace*{0.1em}\raisebox{-0.45ex}{\includegraphics[width=1em]{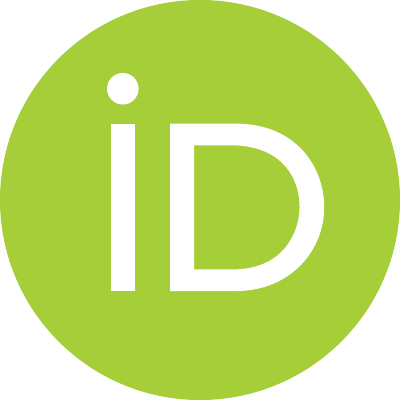}}}}

\usepackage{todonotes}

\begin{document}

\title{\boldmath First Observation of $D^{0(+)}\to \bar K\omega e^+\nu_e$ and Determination of the Branching Fraction of $\bar K_1(1270)\to \bar K \omega$}

\author{
M.~Ablikim$^{1}$\BESIIIorcid{0000-0002-3935-619X},
M.~N.~Achasov$^{4,b}$\BESIIIorcid{0000-0002-9400-8622},
P.~Adlarson$^{80}$\BESIIIorcid{0000-0001-6280-3851},
X.~C.~Ai$^{85}$\BESIIIorcid{0000-0003-3856-2415},
R.~Aliberti$^{37}$\BESIIIorcid{0000-0003-3500-4012},
A.~Amoroso$^{79A,79C}$\BESIIIorcid{0000-0002-3095-8610},
Q.~An$^{76,62,\dagger}$,
Y.~Bai$^{60}$\BESIIIorcid{0000-0001-6593-5665},
O.~Bakina$^{38}$\BESIIIorcid{0009-0005-0719-7461},
Y.~Ban$^{48,g}$\BESIIIorcid{0000-0002-1912-0374}
% ...
\\
\vspace{0.2cm}
(BESIII Collaboration)\\
\vspace{0.2cm} {\it
$^{1}$ Institute of High Energy Physics, Beijing 100049, People's Republic of China\\
% ...
$^{4}$ Budker Institute of Nuclear Physics SB RAS (BINP), Novosibirsk 630090, Russia\\
% ...
$^{37}$ Johannes Gutenberg University of Mainz, Johann-Joachim-Becher-Weg 45, D-55099 Mainz, Germany\\
$^{38}$ Joint Institute for Nuclear Research, 141980 Dubna, Moscow region, Russia\\
% ...
$^{48}$ Peking University, Beijing 100871, People's Republic of China\\
% ...
$^{60}$ Southeast University, Nanjing 211100, People's Republic of China\\
% ...
$^{62}$ State Key Laboratory of Particle Detection and Electronics, Beijing 100049, Hefei 230026, People's Republic of China\\
% ...
$^{76}$ University of Science and Technology of China, Hefei 230026, People's Republic of China\\
% ...
$^{79}$ University of Turin and INFN, (A)University of Turin, I-10125, Turin, Italy; (B)University of Eastern Piedmont, I-15121, Alessandria, Italy; (C)INFN, I-10125, Turin, Italy\\
$^{80}$ Uppsala University, Box 516, SE-75120 Uppsala, Sweden\\
% ...
$^{85}$ Zhengzhou University, Zhengzhou 450001, People's Republic of China\\
\vspace{0.2cm}
$^{\dagger}$ Deceased\\
% ...
$^{b}$ Also at the Novosibirsk State University, Novosibirsk, 630090, Russia\\
% ...
$^{g}$ Also at State Key Laboratory of Nuclear Physics and Technology, Peking University, Beijing 100871, People's Republic of China\\
% ...
}}

\begin{abstract}
  Using 20.3~fb$^{-1}$ of $e^+e^-$ annihilation data collected at a center-of-mass energy of 3.773~GeV with the BESIII detector, we report the first observation of the semileptonic decays \mbox{$D^0\to K^-\omega e^+\nu_e$} and \mbox{$D^+\to K_S^0\omega e^+\nu_e$} with significances of $8.0\sigma$ and $5.8\sigma$, respectively. 
  Their decay branching fractions are measured to be ${\cal B}(D^0\to K^-\omega e^+\nu_e)=(9.4^{+2.0}_{-1.8}\pm 0.6)\times10^{-5}$ and ${\cal B}(D^+\to K_S^0\omega e^+\nu_e)=(8.0^{+2.4}_{-2.1}\pm 0.7)\times10^{-5}$. 
  Combining with the latest measurements of $D^{0(+)}\to  K^-\pi^+\pi^{-(0)} e^+\nu_e$ and assuming $\bar{K}_1(1270)$ to be the sole mediating resonance in all processes, the branching ratios are determined to be
  $\frac{\Gamma(K_1(1270)^-\to K^-\pi^+\pi^-)}{\Gamma(K_1(1270)^-\to K^-\omega)} = 3.4^{+0.8}_{-0.7} \pm 0.3$ and
  $\frac{\Gamma(\bar{K}_1(1270)^0\to K^-\pi^+\pi^0)}{\Gamma(\bar{K}_1(1270)^0\to \bar{K}^0\omega)} = 7.9^{+2.4}_{-2.1} \pm 0.7$.
  The combined branching fraction is determined to be
  $\mathcal B(\bar{K}_1(1270)\to \bar{K}\omega) = (8.4\pm 1.4 \pm 0.5)\%$, which is the most precise measurement from a collider experiment. 
 Furthermore, 
  for the first time based on a single vector-pseudoscalar decay channel, the $\bar{K}_1(1270)$ mass is measured to be $(1336 \pm 9 \pm 2)$~MeV/$c^2$, providing unique experimental support for the two-pole structure of the $\bar{K}_1(1270)$ resonance. 
  The first uncertainties are statistical, and the second are systematic.
\end{abstract}
\maketitle

Semileptonic~(SL) decays of $D^{0(+)}$ mesons offer an ideal testbed to  understand non-perturbative strong-interaction dynamics in weak decays~\cite{SLD1, SLD2}. The SL transitions of $D^{0(+)}$ mesons into the axial-vector meson $\bar{K}_1(1270)$, which provide valuable information on the $\bar{K}_1(1270)$, are especially appealing. Ref.~\cite{gammapolar} notes that the combined measurements of \mbox{$D^{0(+)} \to \bar{K}_1(1270)\ell^+ \nu_\ell$} and \mbox{$\bar{B} \to \bar{K}_1(1270)\gamma$} offer a possible approach for determining the photon polarization in $b \to s \gamma$ transitions with no theoretical ambiguity. Furthermore, improved knowledge of the $\bar{K}_1(1270)$ will facilitate the search for new physics in $\bar{B}\to \bar{K}_1(1270) \, \ell^+\ell^-$, which has recently drawn much interest from both theorists~\cite{Bhatta:2020cyt, Hisaki:2008, Li:2011, Ishaq:2013, Munir:2016, Huang:2019, Munir:2022, Bhutta:2025} and experimentalists~\cite{LHCb:2024yci}. 

The SL decays $D^{0(+)}\to \bar{K}_1(1270)e^+\nu_e$ have been observed in the modes of $\bar{K}_1(1270)^0\to K^-\pi^+\pi^0$, $K_1(1270)^-\to K^-\pi^+\pi^-$, and $\bar{K}_1(1270)^{(0)-}\to K_S^0\pi^-\pi^{+(0)}$~\cite{liuke, Fyl, Tya}. Most recently, the first amplitude and angular analyses of the five-body decays of $D^{0(+)}\to K^-\pi^+\pi^{-(0)}e^+ \nu_e$ have been performed, where the $\bar{K}_1(1270)$ is found to be the only intermediate state in the $\bar{K}\pi\pi$ system~\cite{zyj}. 
As the only notable four-body $\bar{K}_1(1270)$ decay, studies of the $\bar{K}_1(1270) \to \bar{K}\omega$ with $\omega \to \pi^+\pi^-\pi^0$ decay will aid our understanding of the $\bar{K}_1(1270)$ line shape, and provide important external inputs for establishing a more reliable model of $\bar{K}_1(1270) \to \bar{K}\pi\pi$. 
A two-pole interpretation for the $\bar{K}_1(1270)$ meson has been proposed and exploited theoretically~\cite{Roca:2005, Geng:2007, Wang:2020, Dias:2021}, where the lower (higher) mass pole couples predominantly to the $\bar{K}^*\pi$ ($\rho\bar{K}$) channel. Due to the phase-space limitation and narrow $\omega$ width, the lower pole at $\sim1195$~MeV/$c^2$ can hardly couple to the $\bar{K}\omega$ channel, making the $\bar{K}\omega$ mode an ideal probe to isolate the higher mass pole and test the two-pole hypothesis of the $\bar{K}_1(1270)$ meson. 
Ref.~\cite{K1strong} points out that theoretical discussions have often underestimated some difficulties and complications in studying $K_1 \to K\pi\pi$ decays. These three-body processes involve multiple interfering intermediate channels. Further complications arise from the large widths of the $K^*$ and $\rho$ isobars, as well as from the overlap and mixing of the two close $K_1(1270)$ and $K_1(1400)$ states.
The width of the $\omega$ meson is relatively narrow compared with $\rho$ and $\bar{K}^*(892)$, substantially suppressing any bias caused by the modeling of intermediate-state width effects. The dominance of the $\bar{K}\omega$ resonance in the $\bar{K}_1(1270) \to \bar{K}\pi^+\pi^-\pi^0$ channel, which is supported by the result of this analysis, eliminates the multi-channel interference effects prevalent in the $\bar{K}_1(1270) \to \bar{K}\pi\pi$ decays, where significant complexity arises from the interplay of the dominant quasi-two-body pathways such as $\bar{K}_1(1270) \to \bar{K}^*\pi$ and $\bar{K}_1(1270) \to \bar{K}\rho$. Finally, the near absence of the $\bar{K}_1(1400), \bar{K}^*(1410), \bar{K}^*(1430)$, which is consistent with the findings of this analysis, suppresses interference from these states, enabling isolation of the $\bar{K}_1(1270)$ dynamics. 

The current world average of the branching fraction~(BF) of $\bar{K}_1(1270) \to \bar{K}\omega$ is $(11\pm 2)$\%~\cite{PDG}. It comes from only one measurement conducted in a $K^- p\to K^-\pi^-\pi^+ p$ scattering experiment more than forty years ago~\cite{CNTR}, determined indirectly under the assumption that the $\bar{K}\omega$ amplitude is proportional to the $\bar{K}\rho$ amplitude. 
The first measurement in a collider experiment is from the Belle collaboration, reporting ${\cal B}(\bar{K}_1(1270)\to\bar K\omega)=(22.5 \pm 5.2)$\% via an amplitude analysis of $B^+ \to J/\psi K^+\pi^+\pi^-$ 
~\cite{Belle}. 
Recently, the LHCb collaboration has reported ${\cal B}(\bar{K}_1(1270) \to \bar{K}\omega)=(6.3 \pm 3.1)$\% via an amplitude analysis of $B^+ \to \psi(3686)K^+\pi^+\pi^-$ 
~\cite{LHCb}. All these measurements are based on the $K_1(1270)^-\to K^-\pi^+\pi^-$ transition, while no experimental result for $\bar{K}_1(1270)^0\to K_S^0\omega$ is available to date. The $D^{0(+)}\to \bar{K}_1(1270)(\to \bar{K}\omega) e^+\nu_e$ decay with $\omega\to \pi^+\pi^-\pi^0$ provides an independent and clean opportunity to measure ${\cal B}(\bar{K}_1(1270) \to \bar{K}\omega)$ in both neutral and charged channels.
This Letter presents the first measurement of the SL decays $D^0\to K^-\omega e^+\nu_e$ and \mbox{$D^+\to K_S^0\omega e^+\nu_e$} using $20.3\,\mathrm{fb}^{-1}$ of $e^+e^-$ annihilation data collected at a center-of-mass energy of 3.773~GeV with the BESIII detector. Throughout this Letter, the charged-conjugated modes are always implied.

A description of the design and performance of the BESIII detector can be found in Refs.~\cite{BES3,Yu:2016cof}. The inclusive Monte Carlo (MC) sample, described in Refs.~\cite{geant4,kkmc,evtgen,lundcharm,photos}, has been validated to model the background. The exclusive signal samples are generated by simulating the $D^{0(+)}\to \bar{K}_1(1270)e^+\nu_e$ decays with the ISGW2 model, an update of the ISGW quark model for semileptonic meson decays~\cite{SLD2}, and the $\bar{K}_1(1270)$ resonance shape is parameterized by a relativistic Breit-Wigner function with a mass of 1272~MeV/$c^2$ and a width of 90~MeV. The optimization of selection criteria is performed using these initial signal MC samples. The final signal efficiencies and signal shape are determined using updated signal MC samples with a mass of 1336~MeV/$c^2$ measured in this work and a width of 168~MeV~\cite{zyj}. 
The subsequent decay of $\bar{K}_1(1270)\to \bar K\omega $ is described by the VVS\_PWAVE model, which parameterizes the decay of a vector particle to a vector and a scalar particle~\cite{evtgen}. The subsequent decay of $\omega\to\pi^+\pi^-\pi^0$ is modeled using the OMEGA\_DALITZ model that describes the Dalitz amplitude for the decay~\cite{omega2pipipi0}.

At $\sqrt{s}=3.773$~GeV, the $D$ and $\bar{D}$ mesons are produced in pairs without accompanying particles in the final state, which enables the study SL $D$ decays with the double-tag~(DT) method~\cite{DTmethod}. In this method, single-tag (ST) events are those in which $\bar{D}$ mesons are fully reconstructed via the hadronic decay modes \mbox{$\bar{D}^0 \to K^+ \pi^-$}, $K^+ \pi^- \pi^0$, and $K^+ \pi^+ \pi^- \pi^-$; \mbox{$D^- \to K^+ \pi^- \pi^-$}, $K_S^0 \pi^-$, $K^+ \pi^- \pi^- \pi^0$, $K_S^0 \pi^- \pi^0 $, $K_S^0 \pi^+ \pi^- \pi^-$ and $K^+ K^- \pi^-$. In the presence of the ST $\bar{D}$ mesons, candidates for $D^{0(+)}\to K^{-(0)}_{(S)}\omega e^+\nu_e$, $\omega\to\pi^+\pi^-\pi^0$ are selected to form DT events. The BF of the signal decay is determined by
\begin{equation}
\label{eq:BF}
	\mathcal{B}_{\rm sig} = \frac{N_{\rm DT}}{N^{\rm tot}_{\rm ST}\bar{\epsilon}_{\rm sig}},
\end{equation}
where $N^{\rm tot}_{\rm ST}=\sum_{i}{N^i_{\rm ST}}$ and $N_{\rm DT}$ are the total ST and DT yields after summing over all ST modes; $\bar{\epsilon}_{\rm sig}=\sum_{i}\frac{N^i_{\rm ST}}{N^{\rm tot}_{\rm ST}}\frac{\epsilon^i_{\rm DT}}{\epsilon^i_{\rm ST}}$ is the averaged signal efficiency of selecting $D^{0(+)}\to K^{-(0)}_{(S)}\omega e^+\nu_e$ in the presence of the ST $\bar{D}$ meson, where $\epsilon^i_{\rm ST}$ and $\epsilon^i_{\rm DT}$ are the ST and DT efficiencies for the $i^{th}$ ST mode, respectively.

To reconstruct ST $\bar{D}$ mesons, the selection criteria of $\pi^\pm$, $K^\pm$, $K_S^0$, $\gamma$, and $\pi^0$ candidates are the same as in Ref.~\cite{Tya}. Two kinematic variables, the energy difference $\Delta E \equiv E_{\bar{D}}-E_{\rm beam}$ and the beam-constrained mass $M_{\rm BC}\equiv \sqrt{E_{\rm beam}^2-|\vec{p}_{\bar{D}}|^2}$, are used to distinguish the ST $\bar{D}$ mesons from combinatorial backgrounds, where $E_{\rm beam}$ is the beam energy and $(E_{\bar{D}},\vec{p}_{\bar{D}})$ is the four-momentum of the ST $\bar{D}$ meson in the $e^+e^-$ rest frame. The combination with the smallest $|\Delta E|$ is chosen if there are multiple combinations in an event.

The ST candidates are required to satisfy mode-dependent requirements of $\Delta E$, corresponding to about $\pm 3.5\sigma$ around the fitted peaks. The ST yields in data for each ST mode are extracted by fitting individual $M_{\rm BC}$ distributions. Candidates with $M_{\rm BC}\in [1.863,\,1.877]$~GeV/$c^2$ are retained for further analysis. The $\Delta E$ requirements, the ST yields in data, and the ST efficiencies for different ST modes are provided in Ref.~\cite{Supplement}. Summing over all ST modes yields $N^{\rm tot}_{\rm ST} = (16135.4\pm4.6)\times10^3$ for $\bar{D}^0$ and $(10646.9\pm3.8)\times10^3$ for $D^-$.

In the presence of the ST $\bar{D}$ mesons, candidates for \mbox{$D^{0(+)}\to K^{-(0)}_{(S)}\omega e^+\nu_e$} are selected from the remaining tracks and showers. The selection criteria of $K^-,\pi^\pm$ and $\pi^0$ are the same as those used in the ST selection. We require that there is no extra tracks. For the case of multiple $\pi^0$ candidates, the one with the invariant mass closest to the known $\pi^0$ mass is selected~\cite{PDG}. The $\omega$ meson is reconstructed in the $\pi^+\pi^-\pi^0$ decay mode, and its invariant mass $M_{\pi^+\pi^-\pi^0}$ is required to be in the range of [0.60, 0.95]~GeV$/c^2$. 
The selection criteria of $K_S^0$ are the same as those used in the ST selection except for the removal of the $L/\sigma_{L}>2$ requirement, where $L$ is the distance between the common vertex of the $\pi^+\pi^-$ pair and the interaction point, and $\sigma_{L}$ is its uncertainty. For the case of multiple $K_S^0$ candidates, the one with the largest $L/\sigma_{L}$ is chosen.
The $e^+$ candidates are selected by performing particle identification~(PID) based on the specific ionization energy loss (${\rm d}E/{\rm d}x$), time of flight, and electromagnetic calorimeter~(EMC) information. Confidence levels for the positron, pion, and kaon hypotheses are calculated, denoted as $C\!L_{e,\pi,K}$. Charged tracks are assigned as $e^+$ candidates if $C\!L_{e}>0.001$ and $C\!L_{e}/(C\!L_{e}+C\!L_{\pi}+C\!L_{K})>0.8$. The efficiency of this basic PID requirement for $e^+$ candidates is about 95\%. To further distinguish the positron candidates from hadrons, the $e^+$ candidates are required to satisfy $E_e/p_e c>0.67$, $\chi_{{\rm d}E/{\rm d}x}^2<4.9$ for the $D^0$ channel, and $E_e/p_e c>0.66$, $\chi_{{\rm d}E/{\rm d}x}^2<6.0$ for the $D^+$ channel. Here, $E_e$ is the deposited positron energy in the EMC, $p_e$ is the positron momentum measured by the drift chamber~(MDC), and $\chi^2_{{\rm d}E/{\rm d}x}$ is the $\chi^2$ for the positron hypothesis based on the ${\rm d}E/{\rm d}x$ information. The efficiencies of the requirements on $E_e/p_e$ and $\chi_{{\rm d}E/{\rm d}x}^2$ are about 75\% and 79\% for the $D^0$ and $D^+$ channels, respectively. 

Since the neutrino cannot be detected by the BESIII detector, its inferred four-momentum $(E_{\rm miss},\vec{p}_{\rm miss})$ is obtained by calculating the missing energy and momentum via $E_{\rm miss}= E_{\rm beam}-\sum_{j}E_j$ and $\vec{p}_{\rm miss}= \vec{p}_{D}-\sum_j\vec{p}_{j}$. Here, the index $j$ sums over the $K^{-(0)}_{(S)},\pi^+,\pi^-,\pi^0$, and $e^+$ of the signal candidates and $(E_j,\vec{p}_{j})$ is the four-momentum of the $j^{th}$ particle. A kinematic variable $U_{\rm miss}\equiv E_{\rm miss} - |\vec{p}_{\rm miss}|$ is employed to extract the signal yield. To further improve the resolutions of the mentioned variables, the momentum of the $D$ meson is constrained as $\vec{p}_{D}=-\hat{p}_{\bar{D}}\sqrt{E^2_{\rm beam}-m^2_{\bar{D}}}$, where $\hat{p}_{\bar{D}}$ is the unit vector in the direction of the ST $\bar{D}$ meson and $m_{\bar{D}}$ is the known $\bar{D}$ mass~\cite{PDG}.

Further requirements are applied to suppress background. The maximum shower energy for extra photons not used in event reconstruction, $E^{\rm max}_{{\rm extra}-\gamma}$, is required to be less than 0.3~GeV.  The opening angle between $e^+$ and $\pi^-$ from $\omega$ is required to satisfy $\cos\theta_{e^+\pi^-}<0.87$ and $\cos\theta_{e^+\pi^-}<0.92$ for the $D^0$ and $D^+$ channels, respectively, to suppress the background of $D^{0(+)}\to K_{(S)}^{-(0)} \pi^+\pi^0 \pi^0$, with $\pi^0\to e^+e^-\gamma$ for one of the two $\pi^0$s.  
The invariant mass of $K^{-(0)}_{(S)}\pi^+\pi^-\pi^0\pi^+_{e\to\pi}$, $M_{K^{-(0)}_{(S)}\pi^+\pi^-\pi^0\pi^+_{e\to\pi}}$, is required to be less than 1.77~GeV/$c^2$ and 1.80~GeV/$c^2$ for the $D^0$ and $D^+$ channels to reject the hadronic decays $D^{0(+)}\to K^{-(0)}_{(S)}\pi^+\pi^-\pi^+\pi^0$, where $\pi^+_{e\to\pi}$ denotes using pion mass hypothesis for the $e^+$ track. For the $D^+$ channel, the energy of the $\pi^0$ candidate must be greater than 0.17~GeV to suppress the background from misidentified $\pi^0$ candidates.  The opening angle between the missing momentum and the most energetic unused shower is required to satisfy $\cos(\theta_{{\rm miss},\gamma})<0.9$, to suppress background from $D$ hadronic decays with multiple $\pi^0$s and missing a $\pi^0$. Different requirements for the $D^0$ and $D^+$ channels are obtained by optimizing a figure-of-merit defined as $S/\sqrt{S+B}$, where $S$ and $B$ are the signal and background yields according to MC simulation, respectively.

\begin{figure}[htbp]
	\centering
	\setlength{\abovecaptionskip}{0.cm}
	\includegraphics[width=8.7cm]{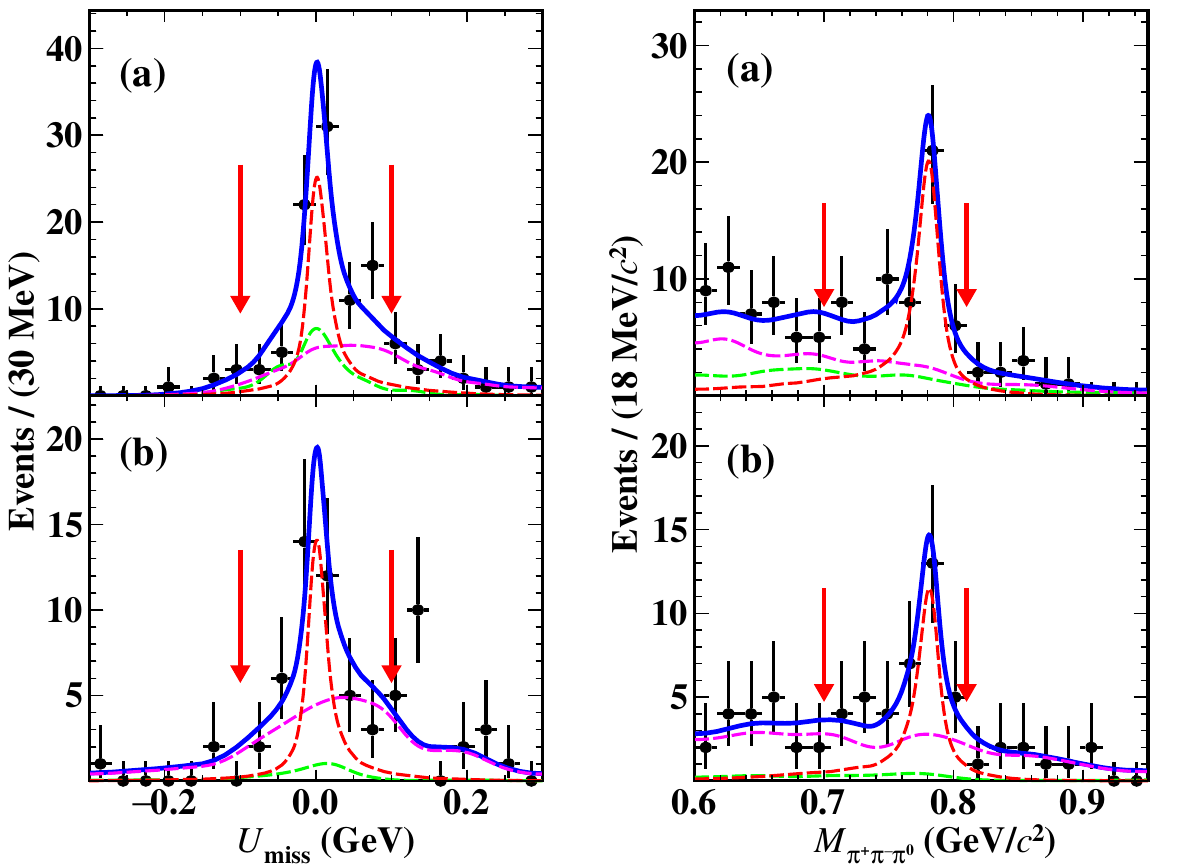}
	\caption{Projections of the fits to the $U_{\rm miss}$ versus $M_{\pi^+\pi^-\pi^0}$ distributions of the (a) $D^0$ and (b) $D^+$ channels. The points with error bars are data. The solid blue lines denote the total fits. The dashed red, green, and magenta lines show the signal, $D^{0(+)}\to K^{-(0)}_{(S)}\pi^+\pi^-\pi^+\pi^0$ peaking background, and non-peaking background contributions, respectively. The red arrows indicate the requirements $|U_{\rm miss}|<0.1$~GeV and $M_{\pi^+\pi^-\pi^0}\in[0.70,0.81]$~GeV$/c^2$.}
	\label{fig:fit}
\end{figure}

To extract the signal yields of $D^{0(+)}\to K^{-(0)}_{(S)}\omega e^+\nu_e$, two-dimensional (2D) unbinned maximum likelihood fits are performed on the $U_{\rm miss}$ versus $M_{\pi^+\pi^-\pi^0}$ distributions of the accepted candidates with $|U_{\rm miss}| < 0.30$ GeV and $0.60 < M_{\pi^+\pi^-\pi^0}<0.95$ GeV, as shown in Fig.~\ref{fig:fit}. $M_{\pi^+\pi^-\pi^0}$ is chosen instead of $M_{K\pi^+\pi^-\pi^0}$ because of the significantly larger width of $\bar{K}_1(1270)$ compared with $\omega$. In each fit, the signal is described by the MC-simulated shape, while both the peaking and non-peaking background shapes are derived from the inclusive MC sample. The smooth
2D probability density functions (PDFs) of signal and background
are modeled using RooNDKeysPdf~\cite{roofit,class}.
The sizes of the $D^{0(+)}\to K^{-(0)}_{(S)}\pi^+\pi^-\pi^+\pi^0$ peaking background contributions are also extracted from the inclusive MC sample. For the $D^0$ channel, the number of $D^0\to K^-\pi^+\pi^-\pi^+\pi^0$ peaking background events is estimated from MC simulation, modified by a correction factor of $1.08 \pm 0.10$, based on the data-MC inconsistency in the vetoed region of $M_{K^{-}\pi^+\pi^-\pi^0\pi^+_{e\to\pi}}>1.77$~GeV/$c^2$. For the $D^+$ channel, the number of $D^+\to K_S^0\pi^+\pi^-\pi^+\pi^0$ peaking background events is fixed according to MC simulation. In this case, no correction is applied due to a much lower peaking contribution. The yields of the signal and non-peaking background components are allowed to float. The signals of $D^0\to K^-\omega e^+\nu_e$ and $D^+\to K_S^0\omega e^+\nu_e$ are observed with significances of $8.0\sigma$ and $5.8\sigma$, respectively. The significance is calculated as $\sqrt{-2\Delta {\rm ln} L}$, where $\Delta {\rm ln} L$ is the difference of the log-likelihoods with and without the signal component in the fit, taking into account systematic uncertainties of the 2D fits. The signal efficiencies are estimated based on the signal MC samples. The fitted signal yields from the data, the average signal efficiencies and the obtained BFs are summarized in Table~\ref{tab:Bfs}.
\begin{table}[htbp]
	\setlength{\abovecaptionskip}{0.0cm}
	\centering
        \renewcommand\arraystretch{1.3}
	\setlength\tabcolsep{5pt}
	\caption{Fitted signal yields in data ($N_{\rm DT}$), averaged signal efficiencies ($\bar{\epsilon}_{\rm sig}$), and the obtained BFs ($\mathcal{B}_{\rm sig}$), where the first uncertainties are statistical and the second systematic.}
    \begin{tabular}{cccc}
		\hline\hline
		Channel& $N_{\rm DT}$ &$\bar{\epsilon}_{\rm sig}$~(\%)&$\mathcal{B}_{\rm sig}~(\times10^{-5})$ \\
		\hline
		$D^0$ & $42.2^{+9.2}_{-8.3}$ & $3.08\pm0.03$ & $9.4^{+2.0}_{-1.8}\pm 0.6$ \\
		$D^+$ & $21.4^{+6.5}_{-5.7}$ & $4.08\pm0.04$ & $8.0^{+2.4}_{-2.1}\pm 0.7$\\
		\hline\hline
	\end{tabular}
	\label{tab:Bfs}
\end{table}

Table~\ref{tab:sys} summarizes the systematic uncertainties in the BF measurement from different sources. The total uncertainties are calculated by combining all individual contributions in quadrature.
\begin{table}[htpb]
\caption{Relative systematic uncertainties, in \%, for the BF measurements.}
\centering
\setlength\tabcolsep{15pt}
\begin{tabular}{ccc}
\hline\hline
     Source & $D^0$ & $D^+$\\
     \hline
     $N_{\rm ST}^{\rm tot}$        & 0.3 & 0.3 \\
     MC statistics                 & 0.3 & 0.3 \\
     $\omega\to\pi^+\pi^-\pi^0$ BF & 0.8 & 0.8 \\
     Tracking                      & 1.0 & 1.8 \\
     PID                           & 2.1 & 1.7 \\
     $\pi^0$ reconstruction        & 1.0 & 1.0 \\
     $K_{S}^{0}$ reconstruction    & -   & 1.0 \\
     Kinematic requirements        & 1.5 & 1.3 \\
     2D fit                        & 1.0 & 2.4 \\
     $\bar{K}_1(1270)$ line shape  & 2.9 & 5.1 \\
     Signal decay model            & 4.9 & 5.4 \\
     \hline
     Total                         & 6.5 & 8.5 \\
     \hline\hline\\
     \label{tab:sys}
\end{tabular}
\end{table}
Details of the systematic uncertainties in the BF measurement are discussed below. The uncertainty associated with the ST yield $N^{\rm tot}_{\rm ST}$ is estimated to be 0.3\% by varying the signal and background shapes~\cite{BESIII:2024zfv}. The systematic uncertainty due to MC statistics is calculated by $\sqrt{\frac{1-\epsilon}{N \, \epsilon}}$ to be 0.3\% for both decays, where $N$ is the number of events generated in the signal MC sample and $\epsilon$ is the signal efficiency. The uncertainties from the quoted BFs of $K_{S}^{0}\to \pi^+\pi^-$ and $\pi^0\to\gamma\gamma$ are negligible, and that from $\omega\to\pi^+\pi^-\pi^0$ is 0.8\%~\cite{PDG}.

The efficiencies of tracking and PID of $\pi^\pm$ and $K^-$, $\pi^0$ and $K_S^0$ reconstructions are studied with hadronic $D\bar{D}$ decays. The tracking and PID efficiencies of $\pi^\pm$ or $K^-$ are re-weighted with the data-MC difference, depending on the $\pi^\pm$ or $K^-$
momentum. The uncertainties 
of $\pi^0$ and $K_S^0$ reconstruction efficiencies are assigned to be 1.0\% for each $\pi^0$ and $K_S^0$, according to the momentum dependent data-MC difference. The efficiencies of tracking and PID for the $e^+$ are investigated with $e^+e^-\to\gamma e^+e^-$. The tracking and PID efficiencies of $e^+$ are weighted with the data-MC difference, depending on both momentum and $\cos\theta$ of $e^+$. 

The uncertainties associated with various kinematic cuts are studied with the control samples of $D^{0}\to K_{S}^{0}\pi^{-}e^{+}\nu_{e}$ and $D^{+}\to K_{S}^{0}\pi^{0}e^{+}\nu_{e}$. The cut efficiency differences between data and MC simulation are found to be $(-2.6\pm1.5)\%$ for $D^0$ and $(-1.3\pm1.3)\%$ for $D^+$. After correcting the BF of $D^0\to K^-\omega e^+\nu_e$ by $-2.6$\% and $D^+\to K_S^0\omega e^+\nu_e$ by $-1.3$\%, the uncertainties from variable requirements are assigned as 1.5\% and 1.3\% for the $D^0$ and $D^+$ channels, respectively.

The uncertainties of the 2D fits are estimated by changing the signal and background shapes. 
Two issues of uncertainty from the signal shape are investigated. First, two independent fits are performed with the original signal shape convolved with a Gaussian function for the $U_{\rm miss}$ or $M_{\pi^+\pi^-\pi^0}$ distribution. The change in the signal yields are negligible. Second, alternative fits are performed with the RooKeysPDF smooth parameters of the signal shapes increased by 50\% from the default values of 1.0. The changes in signal yields, 0.2\% and 0.4\%, are assigned as the systematic uncertainties for the $D^0$ and $D^+$ channels, respectively.
The uncertainty from final state radiation~(FSR) of $e^+$ is estimated by an alternative fit with the signal altered 
by FSR recovery, meaning that neutral showers within 5$^\circ$ of the initial positron direction are merged with the four-momentum of the positron measured by the MDC. The changes of signal yields, 0.2\% and 2.0\% for the $D^0$ and $D^+$ channels, are assigned as the uncertainties.
The uncertainties associated with the non-peaking background shape are estimated by varying the relative fractions of the major background processes by the uncertainties of their BFs. The maximum changes of signal yields, 0.2\% and 0.4\% for the $D^0$ and $D^+$ channels, are assigned as the uncertainties.
The uncertainty associated with the peaking background shape for the $D^0$ channel is estimated by varying the correction factor of the number of $D^0\to K^-\pi^+\pi^-\pi^+\pi^0$ events by $\pm 1\sigma$. For the $D^+$ channel, the uncertainty is estimated by modifying the number of $D^+\to K_S^0\pi^+\pi^-\pi^+\pi^0$ events from MC simulation with a correction factor, $0.72\pm 0.23$, acquired using the same method as in the $D^0$ channel. The maximum changes of signal yields, 0.9\% and 1.2\% for the $D^0$ and $D^+$ channels, are assigned as the related uncertainties.
The uncertainties due to contribution from the non-resonant $D^{0(+)}\to K^{-(0)}_{(S)}\pi^+\pi^-\pi^0 e^+\nu_e$ decays are investigated by adding a non-resonant decay component based on MC simulation in the fit with the related yield floating. The fitted yields are consistent with zero, and the changes in signal yields are negligible. Contributions from non-resonant $\bar{K}\omega$ are also estimated to be negligible.
Contributions from other strange resonances, such as $\bar{K}_1(1400)$ and $\bar{K}_2^*(1430)$, are estimated to yield fewer than one event in each channel based on the known BFs or upper limits on the BF of $D\to \bar{K}_1(1400)/\bar{K}_2^*(1430) e^+ \nu_e$~\cite{zyj,D2Kpi0enu}, and are thus neglected.
The listed uncertainties are then combined in quadrature.    

The uncertainties from the $\bar{K}_1(1270)$ lineshape are estimated by varying the mass of $\bar{K}_1(1270)$ by $\pm 1\sigma$, and changing the width into 146~MeV or 200~MeV and re-measure the signal efficiency. The maximum changes of the signal efficiencies, 2.9\% and 5.1\% for the $D^0$ and $D^+$ channels, are assigned as the uncertainties.

The uncertainties due to the signal decay model are estimated by generating $D^{0(+)}\to K^-\pi^+\pi^{-(0)}e^+\nu_e$ MC events using the amplitude model from Ref.~\cite{zyj}, and applying a gradient-boosted reweighting technique~\cite{GBR} to $D^{0(+)}\to \bar{K}\omega e^+\nu_e$ MC events at the generator and reconstructed levels, using $q^2$, $p_{K_1}$, and $p_e$ as input variables. 
The resulting changes in signal efficiencies, $4.9\%$ and $5.4\%$ for the $D^0$ and $D^+$ channels, are assigned as the uncertainties.

\begin{figure}[htbp]
	\centering
	\setlength{\abovecaptionskip}{0.cm}
	\includegraphics[width=4.2cm]{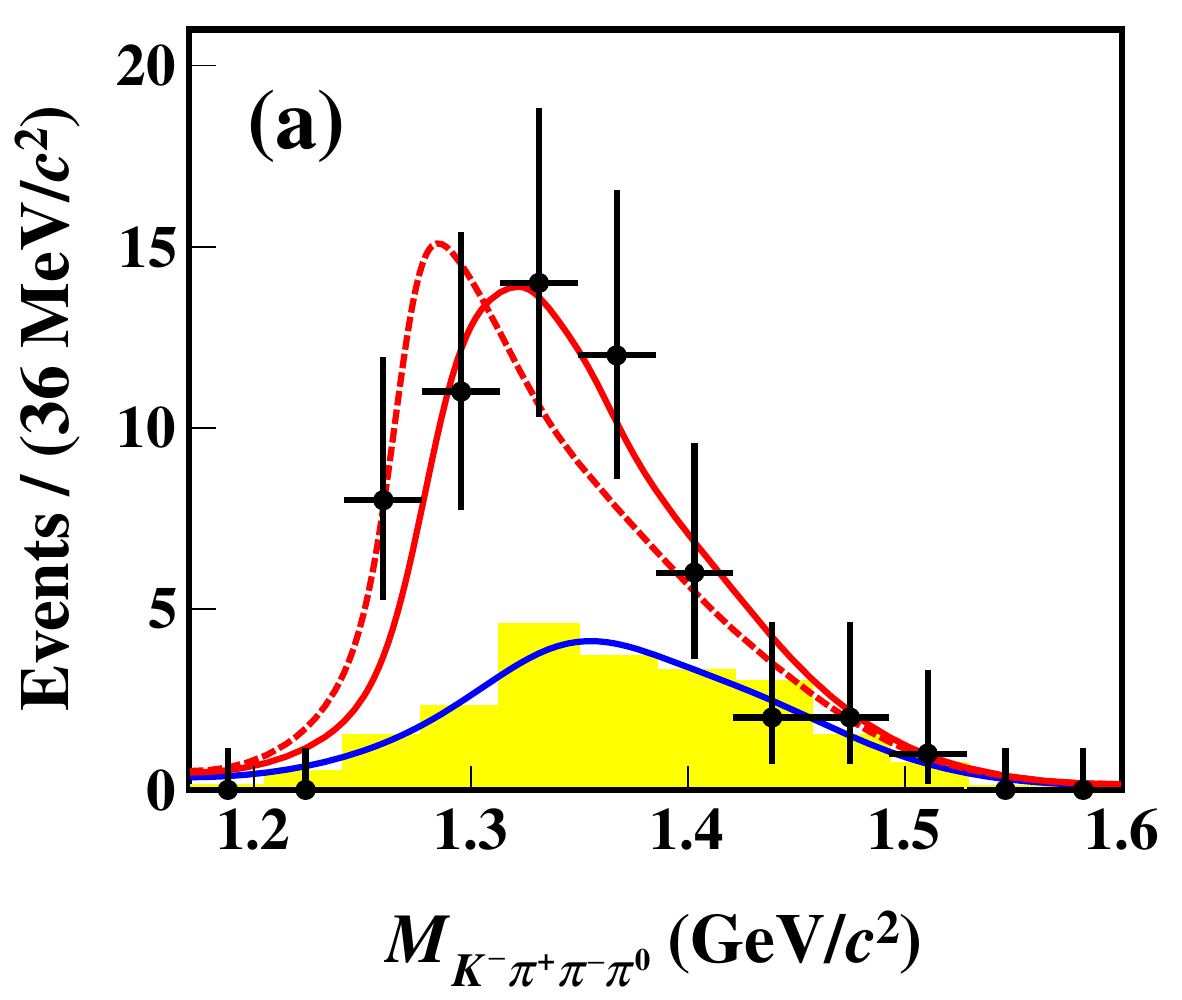}
        \hspace{0.41pt}
        \includegraphics[width=4.2cm]{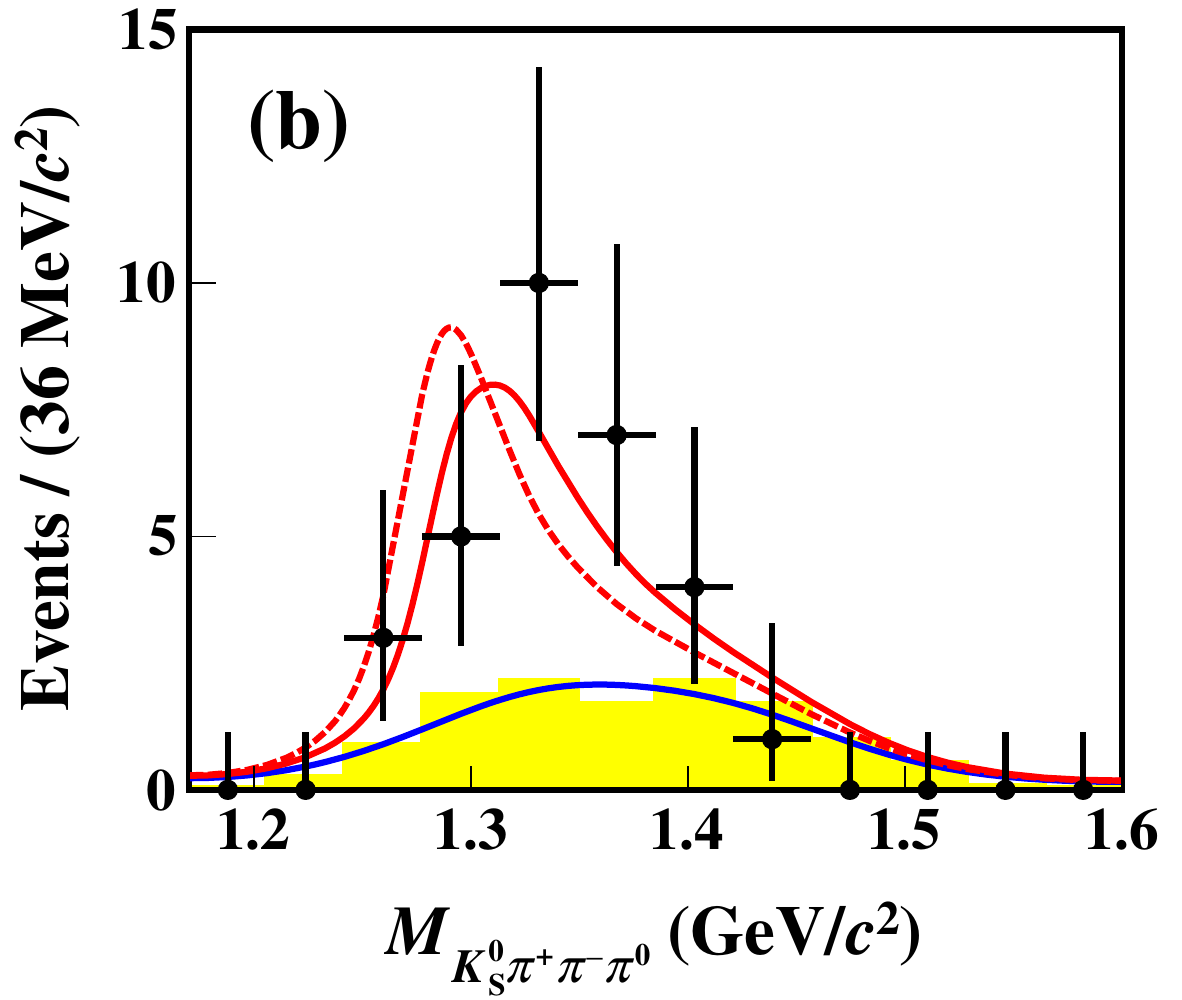}
	\caption{The $M_{\bar{K}\pi^+\pi^-\pi^0}$ distributions of the accepted candidates for the (a) $D^0$ and (b) $D^+$ channels. The points with error bars are data. The yellow histograms and the blue curves denote the estimated contributions and kernel-estimation functions based on the inclusive MC samples, respectively. The red dashed and solid curves denote the total distributions with signal components modeled from signal MC samples with $m_{\rm input}$ set at 1272 and 1336 MeV/$c^2$, respectively. }
	\label{fig:mK1}
\end{figure}

To determine the $\bar{K}_1(1270)$ mass pole, $m^{\bar{K}\omega}_{\bar{K}_1(1270)}$, we perform a simultaneous unbinned maximum-likelihood fit to the $M_{\bar{K}\pi^+\pi^-\pi^0}$ spectra of the $D^0$ and $D^+$ channels in the signal-enhanced region defined by $|U_{\rm miss}|<0.1$~GeV and $M_{\pi^+\pi^-\pi^0}\in[0.70,0.81]$~GeV$/c^2$. 
The modeling of the backgrounds are the same as in the 2D fits. The numbers of both peaking and other background events are fixed to the integrals of the corresponding PDFs in the signal region from the 2D fits, and the signal yields are free to vary. 
In the $D^{0(+)}$ channel, the signal shape is described by the shape from simulated signal events, and convolved with a Gaussian function ${\cal G}(M_{\bar{K}\pi^+\pi^-\pi^0};m_0,\sigma_{D^{0(+)}})$.  
Here $m_0$ is shared between the $D^0$ and $D^+$ channels to measure the pole position, $m^{\bar{K}\omega}_{\bar{K}_1(1270)}=m_{\rm input}+m_0$, with $m_{\rm input}$ being the input $\bar{K}_1(1270)$ mass for the signal MC samples, while the widths $\sigma_{D^{0(+)}}$ are allowed to vary independently. 
The signal MC samples for both channels are iteratively updated with $m_{\rm input}$ shifted by the fitted $m_0$ central value from the prior iteration, until $\left\vert m_0\right\vert$ becomes negligible (within 10\% of its uncertainty). 
We measure $m^{\bar{K}\omega}_{\bar{K}_1(1270)}=(1336 \pm 9 \pm 2)$~MeV$/c^2$, where the first uncertainty is statistical and the second systematic. The systematic uncertainty from signal modeling is estimated by reproducing the signal MC samples based on different input widths of $\bar{K}_1(1270)$, 146~MeV or 200~MeV, and rerunning the simultaneous fit with updated signal lineshapes. The maximum change of the fitted Gaussian mean, 0.5~MeV$/c^2$, is taken as the uncertainty. The systematic uncertainty from background modeling, 1.5~MeV$/c^2$, is estimated by changing background shapes and varying the background yields. To study the uncertainty from possible detector effects, the same strategy is applied to measure the mass pole of $K^*(892)^-$ with a high-statistics control sample of $D^0 \to K^- \pi^0 e^+ \nu_e$. 
The difference of our measurement with the known $K^*(892)^-$ mass~\cite{PDG}, 0.3~MeV$/c^2$, is taken as the uncertainty. The total systematic uncertainty is obtained by summing the individual contributions in quadrature.
Requiring $|U_{\rm miss}|<0.1$~GeV and $M_{\pi^+\pi^-\pi^0}\in [0.70, 0.81]$~GeV$/c^2$
, the $M_{\bar{K}\pi^+\pi^-\pi^0}$ distributions of the accepted candidates for $D^{0(+)}\to K^{-(0)}_{(S)}\omega e^+\nu_e$ are shown in Fig.~\ref{fig:mK1}. The data distributions are consistent with MC predictions that include only signals from $D^{0(+)}\to \bar K_1(1270) e^+\nu_e$.
The final signal MC samples with the input $\bar K_1(1270)$ mass determined by this analysis offer significantly improved description of the data $K_1(1270)$ lineshapes than those from signal MC samples generated with the $K_1(1270)$ mass set at 1272 MeV/$c^2$. 

Assuming the $\bar{K}_1(1270)$ to be the sole mediating resonance in the $D^{0(+)}\to \bar{K}\pi\pi e^+\nu_e$ and $D^{0(+)}\to \bar{K}\omega e^+\nu_e$ decays, the branching ratios are determined to be $\gamma_{D^0} \equiv \frac{\Gamma(K_1(1270)^-\to K^-\pi^+\pi^-)}{\Gamma(K_1(1270)^-\to K^-\omega)} = \frac{\mathcal{B}(D^0\to K^-\pi^+\pi^- e^+\nu_e)}{\mathcal{B}(D^0\to K^-\omega e^+\nu_e)} = 3.4^{+0.8}_{-0.7} \pm 0.3$ and $\gamma_{D^+} \equiv \frac{\Gamma(\bar{K}_1(1270)^0\to K^-\pi^+\pi^0)}{\Gamma(\bar{K}_1(1270)^0\to \bar{K}^0\omega)} = \frac{\mathcal{B}(D^+\to K^-\pi^+\pi^0 e^+\nu_e)}{\mathcal{B}(D^+\to K_S^0\omega e^+\nu_e)} = 7.9^{+2.4}_{-2.1} \pm 0.7$, where the BFs of $D^{0(+)}\to K^-\pi^+\pi^{-(0)} e^+\nu_e$ are taken from Ref.~\cite{zyj}.
Following the formalism in Ref.~\cite{zyj}, by applying isospin relations and neglecting the insignificant contributions from decays other than $\bar{K}_1(1270) \to K^*(892)\pi$, $K \rho$, and $K \omega$, the BF of $K_1(1270)^-\to K^-\pi^+\pi^-$ and $\bar{K}_1(1270)^0\to K^-\pi^+\pi^0$ are written as $\mathcal{B}(K_1(1270)^-\to K^-\pi^+\pi^-) = \frac{3+4\alpha}{9(1+\alpha)} \, \mathcal{B}(\bar{K}_1(1270) \to \bar{K}\pi\pi)$ and $\mathcal{B}(\bar{K}_1(1270)^0\to K^-\pi^+\pi^0) = \frac{6+4\alpha}{9(1+\alpha)} \, \mathcal{B}(\bar{K}_1(1270) \to \bar{K}\pi\pi)$. Here, $\mathcal{B}(\bar{K}_1(1270) \to \bar{K}\pi\pi)=1 - \mathcal{B}(\bar{K}_1(1270)\to \bar{K}\omega)$ is the BF of $\bar{K}_1(1270)$ decay into $\bar{K}\pi\pi$ final states and $\alpha = \frac{\mathcal{B}(\bar{K}_1(1270)\to \bar{K}^*(892)\pi)}{\mathcal{B}(\bar{K}_1(1270)\to \bar{K}\rho)} = (20.3 \pm 2.1 \pm 8.7)\%$~\cite{zyj}. 
Finally, we have $\mathcal{B}(K_1(1270)^-\to K^-\omega) = \left[(\gamma_{D^0}) \, \frac{9(1+\alpha)}{3+4\alpha} +1 \right]^{-1}= (9.4 \pm 1.9 \pm 0.7)\%$ and $\mathcal{B}(\bar{K}_1(1270)^0\to \bar{K}^0\omega) = \left[(\gamma_{D^+}) \, \frac{9(1+\alpha)}{6+4\alpha} +1 \right]^{-1} = (7.3 \pm 2.1 \pm 0.6)\%$. 

\begin{figure}[htbp]
    \centering
    \setlength{\abovecaptionskip}{0.cm}
    \includegraphics[width=8.6cm]{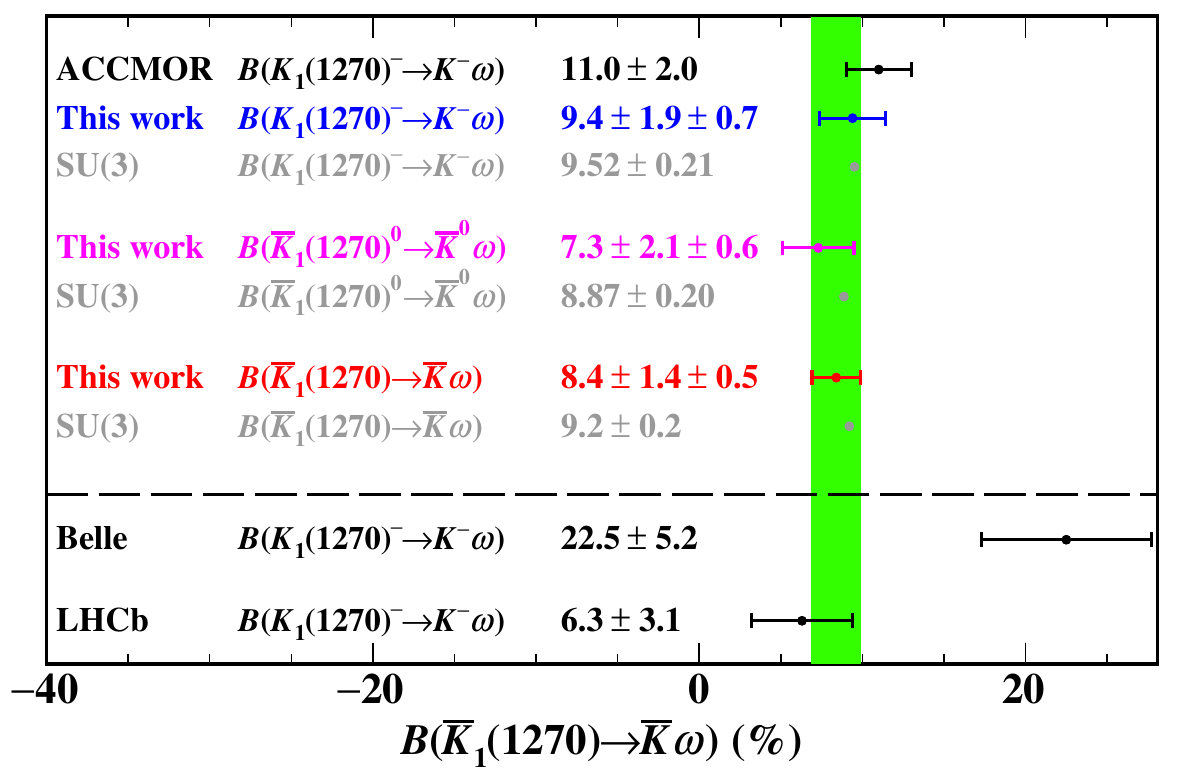}
    \caption{Comparison of the BF of $\bar{K}_1(1270)\to \bar{K}\omega$ obtained in this work with the previous measurements from ACCMOR~\cite{CNTR}, Belle~\cite{Belle}, and LHCb~\cite{LHCb}, as well as the prediction from SU(3) flavor analysis~\cite{SU3}. The green band shows our combined $\mathcal B(\bar{K}_1(1270)\to \bar{K}\omega)$. The uncertainties of the Belle and LHCb measurements are statistical only.}
    \label{fig:rescmp}
\end{figure}

In summary, using $20.3\,\mathrm{fb}^{-1}$ of $e^+e^-$ collision data taken at $\sqrt{s}=$ 3.773~GeV with the BESIII detector, we report the first observation of \mbox{$D^0\to K^-\omega e^+\nu_e$} and \mbox{$D^+\to K_S^0\omega e^+\nu_e$} with significances of $8.0\sigma$ and $5.8\sigma$, respectively, after accounting for systematic uncertainties. Their decay BFs are measured to be ${\cal B}(D^0\to K^-\omega e^+\nu_e)=(9.4^{+2.0}_{-1.8}\pm 0.6)\times10^{-5}$ and ${\cal B}(D^+\to K_S^0\omega e^+\nu_e)=(8.0^{+2.4}_{-2.1}\pm 0.7)\times10^{-5}$. These decays are consistent with being mediated through a sole $\bar{K}_1(1270)$ resonance.
We 
determine ${\mathcal B}(K_1(1270)^-\to K^-\omega) = (9.4 \pm 1.9 \pm 0.7)\%$ and
${\mathcal B}(\bar{K}_1(1270)^0\to \bar{K}^0\omega) = (7.3 \pm 2.1 \pm 0.6)\%$. The $\bar{K}_1(1270)^0\to \bar{K}^0\omega$ decay is observed for the first time.
Under the assumption of isospin symmetry, the combined BF of $\bar{K}_1(1270)\to \bar{K}\omega$ is determined to be $(8.4\pm 1.4 \pm 0.5)\%$. Correlations among the systematic uncertainties are neglected as the total uncertainties are dominated by the statistical components. 
Our combined $\mathcal B(\bar{K}_1(1270)\to \bar{K}\omega)$ is consistent with the only measurement~\cite{CNTR} quoted by the PDG with comparable precision, and consistent with the prediction from SU(3) flavor analysis~\cite{SU3}. 
The comparison of our result with the previous measurements and the theoretical prediction is shown in Fig.~\ref{fig:rescmp}. 
Furthermore, we determine $m^{\bar{K}\omega}_{\bar{K}_1(1270)}=(1336 \pm 9 \pm 2)$~MeV/$c^2$, which is significantly higher than $m^{\bar{K}\pi\pi}_{\bar{K}_1(1270)}=(1271 \pm 3 \pm 7)$~MeV/$c^2$ from $\bar{K}_1(1270)\to \bar{K}\pi\pi$ decays~\cite{zyj}. This difference is naturally expected in the two-pole scenario: the $\bar{K}\pi\pi$ system receives contributions from both mass poles, while the $\bar{K}\omega$ channel predominantly isolates the higher pole. The measured value is also larger than the theoretical higher pole of $\sim1284$~MeV/$c^2$~\cite{Dias:2021}, indicating that the $\bar{K}\omega$ channel may probe the higher pole with additional dynamical effects not fully captured in current models. Our result thus provides unique experimental support for the two-pole structure of the $\bar{K}_1(1270)$ resonance and offers valuable input for refining its theoretical description.
This work is expected to inspire future $\bar{K}_1(1270)$ related studies, in decays such as $\bar B\to \bar{K}\omega \gamma$ and $\bar B\to \bar{K}\omega \ell^+\ell^-$.

The BESIII Collaboration thanks the staff of BEPCII (https://cstr.cn/31109.02.BEPC) and the IHEP computing center for their strong support. This work is supported in part by National Key R\&D Program of China under Contracts Nos. 2023YFA1606000, 2023YFA1606704; National Natural Science Foundation of China (NSFC) under Contracts Nos. 11635010, 11935015, 11935016, 11935018, 12025502, 12035009, 12035013, 12061131003, 12192260, 12192261, 12192262, 12192263, 12192264, 12192265, 12221005, 12225509, 12235017, 12361141819, 12575102; the Chinese Academy of Sciences (CAS) Large-Scale Scientific Facility Program; the Strategic Priority Research Program of Chinese Academy of Sciences under Contract No. XDA0480600; CAS under Contract No. YSBR-101; 100 Talents Program of CAS; The Institute of Nuclear and Particle Physics (INPAC) and Shanghai Key Laboratory for Particle Physics and Cosmology; ERC under Contract No. 758462; German Research Foundation DFG under Contract No. FOR5327; Istituto Nazionale di Fisica Nucleare, Italy; Knut and Alice Wallenberg Foundation under Contracts Nos. 2021.0174, 2021.0299; Ministry of Development of Turkey under Contract No. DPT2006K-120470; National Research Foundation of Korea under Contract No. NRF-2022R1A2C1092335; National Science and Technology fund of Mongolia; Polish National Science Centre under Contract No. 2024/53/B/ST2/00975; STFC (United Kingdom); Swedish Research Council under Contract No. 2019.04595; U. S. Department of Energy under Contract No. DE-FG02-05ER41374


\begin{thebibliography}{99}
    \bibitem{SLD1} N.~Isgur, D.~Scora, B.~Grinstein, and M.~B.~Wise,
    \href{https://doi.org/10.1103/PhysRevD.39.799}
    {Phys. Rev. D \textbf{39}, 799 (1989).}
    
    \bibitem{SLD2} D.~Scora and N.~Isgur,
    \href{https://doi.org/10.1103/PhysRevD.52.2783}
    {Phys. Rev. D \textbf{52}, 2783 (1995).}
    
    \bibitem{gammapolar} W.~Wang, F.~S.~Yu, and Z.X.~Zhao,
    \href{https://doi.org/10.1103/PhysRevLett.125.051802}
    {Phys. Rev. Lett. \textbf{125}, 051802 (2020).}

    \bibitem{Bhatta:2020cyt}
    A.~Bhatta and R.~Mohanta,
    \href{https://doi.org/10.1088/1361-6471/ac012a}
    {J. Phys. G \textbf{48}, 085011 (2021).}

    \bibitem{Hisaki:2008}
    H.~Hatanaka and K.~C.~Yang,
    \href{https://doi.org/10.1103/PhysRevD.78.074007}
    {Phys. Rev. D \textbf{78}, 074007 (2008).}

    \bibitem{Li:2011}
    Y.~Li, J.~Hua, and K.~C.~Yang,
    \href{https://doi.org/10.1140/epjc/s10052-011-1775-2}
    {Eur. Phys. J. C \textbf{71}, 1775 (2011).}

    \bibitem{Ishaq:2013}
    S.~Ishaq, F.~Munir, and I.~Ahmed,
    \href{https://doi.org/10.1007/JHEP07(2013)006}
    {JHEP \textbf{07}, 006 (2013).}

    \bibitem{Munir:2016}
     F.~Munir, S.~Ishaq, and I.~Ahmed,
    \href{https://doi.org/10.1093/ptep/ptv174}
    {Prog. Theor. Exp. Phys. \textbf{2016}, 013B02 (2016).}

    \bibitem{Huang:2019}
    Z.~R.~Huang, M.~A.~Paracha, I.~Ahmed, and C.~D.~Lü,
    \href{https://doi.org/10.1103/PhysRevD.100.055038}
    {Phys. Rev. D \textbf{100}, 055038 (2019).}

    \bibitem{Munir:2022}
    F.~M.~Bhutta, Z.~R.~Huang, C.~D.~Lü, M.~A.~Paracha, and W.~Wang, 
    \href{https://doi.org/10.1016/j.nuclphysb.2022.115763}
    { Nucl. Phys. B \textbf{979}, 115763 (2022).}

    \bibitem{Bhutta:2025}
    F.~M.~Bhutta, A.~Rehman, M.~J.~Aslam, I.~Ahmed, and S.~Ishaq,
    \href{https://doi.org/10.1103/PhysRevD.111.095011}
    {Phys. Rev. D \textbf{111}, 095011 (2025).}
    
    \bibitem{LHCb:2024yci}
    R.~Aaij \textit{et al.} (LHCb Collaboration),
    \href{https://doi.org/10.1103/PhysRevLett.134.181803}
    {Phys. Rev. Lett. \textbf{134}, 181803 (2025).}
    
    \bibitem{liuke} M.~Ablikim {\it et al.} (BESIII Collaboration), 
    \href{https://doi.org/10.1103/PhysRevLett.123.231801}
    {Phys. Rev. Lett. \textbf{123}, 231801 (2019).}
    
    \bibitem{Fyl} M.~Ablikim {\it et al.} (BESIII Collaboration),
    \href{https://doi.org/10.1103/PhysRevLett.127.131801}
    {Phys. Rev. Lett. \textbf{127}, 131801 (2021).}
    
    \bibitem{Tya} M.~Ablikim {\it et al.} (BESIII Collaboration),
    \href{https://doi.org/10.1007/JHEP09(2024)089}
    {JHEP \textbf{09}, 089 (2024).}

    \bibitem{zyj} M.~Ablikim {\it et al.} (BESIII Collaboration),
    \href{https://doi.org/10.1103/xj42-xgzf}
    {Phys. Rev. Lett. \textbf{135}, 091801 (2025).}

    \bibitem{Roca:2005}
    L.~Roca, E.~Oset, J.~Singh,
    \href{https://doi.org/10.1103/PhysRevD.72.014002}
    {Phys. Rev. D \textbf{72}, 014002 (2005).} 
    
    \bibitem{Geng:2007}
    L.~S.~Geng, E.~Oset, L.~Roca and J.~A.~Oller,
    \href{https://doi.org/10.1103/PhysRevD.75.014017}
    {Phys. Rev. D \textbf{75}, 014017 (2007).}   

    \bibitem{Wang:2020}
    G.~Y.~Wang, L.~Roca, and E.~Oset,
    \href{https://doi.org/10.1140/epjc/s10052-020-7939-1}
    {Eur. Phys. J. C \textbf{80}, 388 (2020).}

    \bibitem{Dias:2021}
    J.~M.~Dias, G.~Toledo, L.~Roca, and E.~Oset,
    \href{https://doi.org/10.1103/PhysRevD.103.116019}
    {Phys. Rev. D \textbf{103}, 116019 (2021).}

    \bibitem{K1strong} A.~Tayduganov, E.~Kou, and A.~Le~Yaouanc,
    \href{https://doi.org/10.1103/PhysRevD.85.074011}
    {Phys. Rev. D \textbf{85}, 074011 (2012).}

    \bibitem{PDG} S. Navas {\it et al.} (Particle Data Group),
    \href{https://doi.org/10.1103/PhysRevD.110.030001}
    {Phys. Rev. D \textbf{110}, 030001 (2024).}

    \bibitem{CNTR} C. Daum {\it et al.} (ACCMOR Collaboration), 
    \href{https://doi.org/10.1016/0550-3213(81)90114-0}
    {Nucl. Phys. B \textbf{187}, 1 (1981).}

    \bibitem{Belle} H. Guler {\it et al.} (Belle Collaboration), 
    \href{https://doi.org/10.1103/PhysRevD.83.032005}
    {Phys. Rev. D \textbf{83}, 032005 (2011).}

    \bibitem{LHCb} R.~Aaij {\it et al.} (LHCb Collaboration), 
    \href{https://doi.org/10.1007/JHEP01(2025)054}{JHEP \textbf{01}, 054 (2025).}

    \bibitem{BES3} M. Ablikim {\it et al}. (BESIII Collaboration),
    \href{https://doi.org/10.1016/j.nima.2009.12.050}
    {Nucl. Instrum. Meth. A \textbf{614}, 345 (2010).}
    
    \bibitem{Yu:2016cof} C.~H.~Yu {\it et al.},
    \href{https://doi.org/10.18429/JACoW-IPAC2016-TUYA01}
    {Proceedings of IPAC2016, Busan, Korea, 2016}.

    \bibitem{geant4} S. Agostinelli {\it et al.} (GEANT4 Collaboration),
    \href{https://doi.org/10.1016/S0168-9002(03)01368-8}
    {Nucl. Instrum. Meth. A \textbf{506}, 250 (2003).}

    \bibitem{kkmc}
    S. Jadach, B. F. L. Ward, and Z. W\c{a}s,
    \href{https://doi.org/10.1016/S0010-4655(00)00048-5}
    {Comp. Phys. Commu. \textbf{130}, 260 (2000)};
    \href{https://doi.org/10.1103/PhysRevD.63.113009}
    {Phys. Rev. D \textbf{63}, 113009 (2001).}

    \bibitem{evtgen} D.~J.~Lange,
    \href{https://doi.org/10.1016/S0168-9002(01)00089-4}
    {Nucl. Instrum. Meth. A \textbf{462}, 152 (2001)};
    R.~G.~Ping,
    \href{https://doi.org/10.1088/1674-1137/32/8/001}
    {Chin. Phys. C \textbf{32}, 599 (2008).}

    \bibitem{lundcharm} J. C. Chen, G. S. Huang, X. R. Qi, D. H. Zhang, and Y. S. Zhu,
    \href{https://doi.org/10.1103/PhysRevD.62.034003}
    {Phys. Rev. D \textbf{62}, 034003 (2000).}

    \bibitem{photos} E.~Richter-W\c{a}s,
    \href{https://doi.org/10.1016/0370-2693(93)90062-M}
    {Phys. Lett. B \textbf{303}, 163 (1993).}

    \bibitem{omega2pipipi0} M.~Ablikim {\it et al.} (BESIII Collaboration),
    \href{https://doi.org/10.1103/PhysRevD.98.112007}
    {Phys. Rev. D \textbf{98}, 112007 (2018).}

    \bibitem{DTmethod} R. M. Baltrusaitis {\it et al.} (Mark III Collaboration), 
    \href{https://doi.org/10.1103/PhysRevLett.56.2140}
    {Phys. Rev. Lett. \textbf{56}, 2140 (1986);}
    J. Adler {\it et al.} (Mark III Collaboration), 
    \href{https://doi.org/10.1103/PhysRevLett.60.89}
    {Phys. Rev. Lett. \textbf{60}, 89 (1988).}

    \bibitem{Supplement} See Supplemental Material at [URL will be inserted by publisher] for ST yields and efficiencies.

    \bibitem{roofit} 
    W. Verkerke and D. Kirkby,  eConf No. C0303241, MOLT007 (2003),
    \href{https://arxiv.org/abs/physics/0306116}{arXiv:physics/0306116.}

    \bibitem{class}
    \href{https://root.cern.ch/doc/master/classRooNDKeysPdf.html}{https://root.cern.ch/doc/master/classRooNDKeysPdf.html.}
    
    \bibitem{BESIII:2024zfv} M.~Ablikim \textit{et al.} (BESIII Collaboration),
    \href{https://doi.org/10.1007/JHEP12(2024)206}
    {JHEP \textbf{12}, 206 (2024)}.

    \bibitem{D2Kpi0enu} M.~Ablikim \textit{et al.} (BESIII Collaboration),
    \href{https://arxiv.org/abs/2603.00743}
    {arXiv:2603.00743 [hep-ex].}
    
    \bibitem{GBR}
    \href{https://arogozhnikov.github.io/hep_ml/reweight.html}{https://arogozhnikov.github.io/hep\_ml/reweight.html.}
    
    \bibitem{SU3} Y.~Qiao, Y.~X.~Liu, Y.~G.~Xu, and R.~M.~Wang, 
    \href{https://doi.org/10.1140/epjc/s10052-024-13387-0}
    {Eur. Phys. J. C \textbf{84}, 1110 (2024).}

\end{thebibliography}
\end{document}